\documentclass[]{aa} 

\newcommand{\msun}{$\mathrm{M_{\odot}}$}
\newcommand{\teff}{$T_{\rm eff}$}

\usepackage{color}
\usepackage{soul}
\usepackage{threeparttable}

\usepackage{txfonts}
\usepackage{graphicx}
\usepackage[authoryear]{natbib}
\usepackage{epsfig}

\usepackage{natbib}

\bibpunct{(}{)}{;}{a}{}{,} 

\begin{document}

\title{Solar twins in M67 : Evolutionary status and lithium abundance}
 \subtitle{}

   \author{M. Castro\inst{1},
           J.D. do Nascimento Jr.\inst{1},
           K. Biazzo\inst{2},
           J. Mel\'endez\inst{3,4},
           J.R. De Medeiros\inst{1}
}

   \offprints{M. Castro, email: mcastro@dfte.ufrn.br}

\institute{Departamento de F\'isica Te\'orica e Experimental, Universidade Federal do Rio Grande do Norte, CEP: 59072-970 Natal, RN, Brazil 
\and Istituto Nazionale di Astrofisica, Osservatorio Astrofisico di Arcetri, Largo E. Fermi 5, 50125 Firenze, Italy
\and Centro de Astrof\'isica da Universidade do Porto, Rua das Estrelas, 4150-762 Porto, Portugal
\and Departamento de Astronomia do IAG/USP, Universidade de S\~ao Paulo, Rua do Mat\~ao 1226, S\~ao Paulo, 05508-900, SP, Brazil 
}

\date{Received 05/25/10: Accepted 10/26/10}

\abstract
{}
{We determine the age and mass of the three best solar twin candidates in open cluster M67 through lithium evolutionary models.}
{We computed a grid of evolutionary models with non-standard mixing at metallicity $[Fe/H] = 0.01$ with the Toulouse-Geneva evolution code for a range of stellar masses. We estimated the mass and age of 10 solar analogs belonging to the open cluster M67. We made a detailed study  of the three solar twins of the sample, YPB637, YPB1194, and YPB1787.}
{We obtained a very accurate estimation of the mass of our solar analogs in M67 by interpolating in the grid of evolutionary models. The three solar twins allowed us to estimate the age of the open cluster, which is $3.87^{+0.55}_{-0.66}$ Gyr, which is better constrained than former estimates.}
{Our results show that the 3 solar twin candidates have one solar mass within the errors and that M67 has a solar age within the errors, validating its use as a solar proxy. M67 is an important cluster when searching for solar twins.}
\keywords{Stars: fundamental parameters -- Stars: abundances -- Stars: evolution -- Stars: interiors -- Stars: solar-type}
\titlerunning{Modeling of solar twins in M67}
\authorrunning{Castro et al.}

\maketitle

\section{Introduction}
\label{sec:Intro}

Lithium depletion of each star depends on the evolution of its rotational history and the changes in the convective envelope due to internal mixing. In main-sequence solar-like stars, lithium is easily destroyed by nuclear burning in stellar interiors at temperatures above $2.4 \times 10^6 K$, and its surface abundance indicates the depth of mixing below their photospheres. The Sun has long been thought to be highly Li-depleted by a factor of 10 compared to the other field G stars \citep{lambert&reddy04}. However, two of the best solar twins found by \citet{melendez&ramirez07} (namely, HIP 56948 and HIP 73815) are Li-poor. The amount of Li depletion in solar twins is sensitive to microscopic diffusion, and an extra-mixing process is required to explain the low Li abundances observed, indicating that they also share a similar mixing history with the Sun. 

M67 is one of the most studied open clusters. Its chemical composition, metallicity and age are similar to those of the Sun. Various chemical analyses \citep{tautvaisiene00,randich06,pace08} show that not only abundances of Fe, but also element abundances of O, Na, Mg, Al, Si, Ca, Ti, Cr, and Ni, are very similar to their solar counterparts, as close as allowed by the precision of the measurements. The ages determined are similar to those of the Sun, between 3.5 and 4.8 Gyr \citep{yadav08}. Another interesting aspect of open cluster M67 is its Li-depleted G stars. Lithium abundance study is a crucial point for understanding the physics involved in stellar interiors. \citet{pasquini97} show there is a large spread in Li abundances among solar-type stars in M67. Almost 40\% of their sample shows Li depletion comparable to that of the Sun. In this context, M67 is a perfect target to search for solar analogs and solar twins. Indeed, many stars are similar to the Sun in terms of effective temperature and chemical composition, but their Li abundance is much higher than in our star. Furthermore, among the 90 MS stars studied by \citet{pasquini08} in M67, 10 had similar effective temperature and Li content as low as that of the Sun, making them excellent solar twin candidates.

\begin{figure}[t!]
\begin{center}
\vspace{-0.1in}
\includegraphics[angle=0,height=9cm,width=9cm]{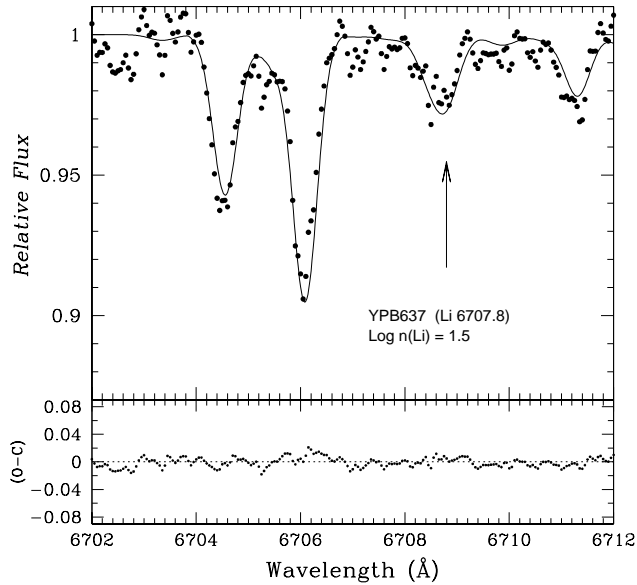}
\end{center}
\vspace{-0.2in}
\caption{Spectral synthesis of the $\lambda  6707$ Li I line of YPB637} 
\label{fig:synt_obj637}
\end{figure}

In \citet{donascimento09}, we showed that it is possible to precisely determine the mass of solar twins using evolutionary models. We determined the mass and the age of 5 solar twins taken from two sources: \citet{melendez&ramirez07} and \citet{takeda07}. We began to compute a solar model, helioseismically calibrated, to reproduce not only solar luminosity and radius, but also Li depletion. To that end, we introduced mixing due to meridional circulation with a feedback effect from the $\mu$-currents \citep{zahn92,vauclair&theado03,theado&vauclair03}. This calibration was used to compute models of different masses whose tracks in a 
\teff-$\log N{\rm (Li)}$ diagram passed through the observational point. In the range of effective temperatures studied (typically 5400-6000 K), Li destruction is very sensitive to mass, and we were able to precisely determine the mass of each solar twin and its age.

\begin{figure}[t!]
\begin{center}
\vspace{-0.1in}
\includegraphics[angle=0,height=9cm,width=9cm]{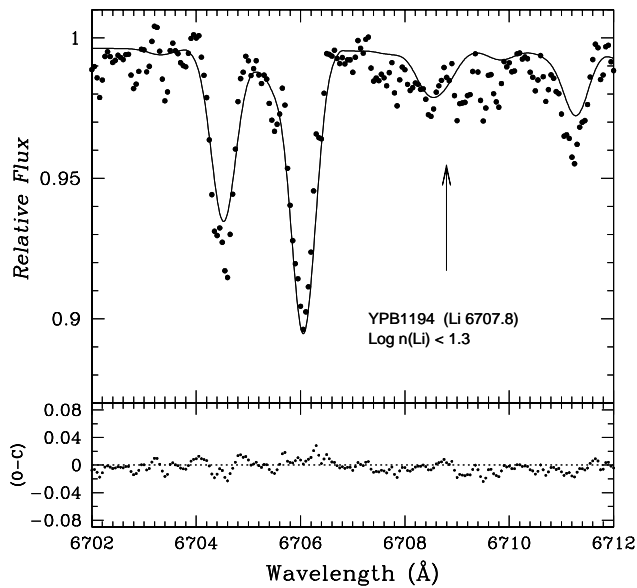}
\end{center}
\vspace{-0.2in}
\caption{Same as Figure \ref{fig:synt_obj637} for YPB1194} 
\label{fig:synt_obj1194}
\end{figure}

In the present work, we computed helioseismically calibrated, evolutionary, solar-like models with microscopic diffusion and rotation-induced mixing in the radiative interior, as in \citet{donascimento09}. We estimate the masses of the 10 solar analogs in M67 found by \citet{pasquini08}. We estimate the age of the three most probable solar twins of the sample. The working sample is described in Sect. \ref{sec:Obs}. Improved Li abundances, using spectral synthesis instead of calibration based on equivalent width (EW, hereafter), are presented in Sect. \ref{sec:obs_lithium}. In Sect. \ref{sec:Models}, we describe the grid of evolutionary models with the non-standard physics that we computed. In Sect. \ref{sec:Results}, we present our results and compare our estimation of the age of M67 with earlier estimate. In this section we also show that the constitutive physics of the models influence and alter the isochrones constructed. We present the M67 color-magnitude diagram constrained by main sequence solar twins. Finally, our conclusions are outlined in Sect. \ref{sec:Conclusions}.

\section{Sample of solar analogs in M67}
\label{sec:Obs}

\citet{pasquini08} observed 90 targets belonging to cluster M67 with the multi-object FLAMES/GIRAFFE spectrograph (see Sect. \ref{sec:obs_lithium}). They computed the effective stellar temperatures of these stars using two spectroscopic methods: the line-depth ratios (LDR) and the H$\alpha$ wings. They derived Li abundances from measured EW using the calibration of \citet{soderblom93}. The authors highlight 10 stars in their sample, for which both $T_{\rm eff}^{\rm LDR}$ and $T_{\rm eff}^{\rm H\alpha}$ are within 100 K of the solar values and which have a strong Li depletion. These stars are good solar twin candidates. However, as observed by the authors, given the limited resolution (R$\sim$17000) and S/N (80-110/pixel in average), the errors in Li abundance determination are high and the stars with upper limits may have a Li abundance comparable to the solar value, or even higher.

\begin{figure}[t!]
\begin{center}
\vspace{-0.1in}
\includegraphics[angle=0,height=9cm,width=9cm]{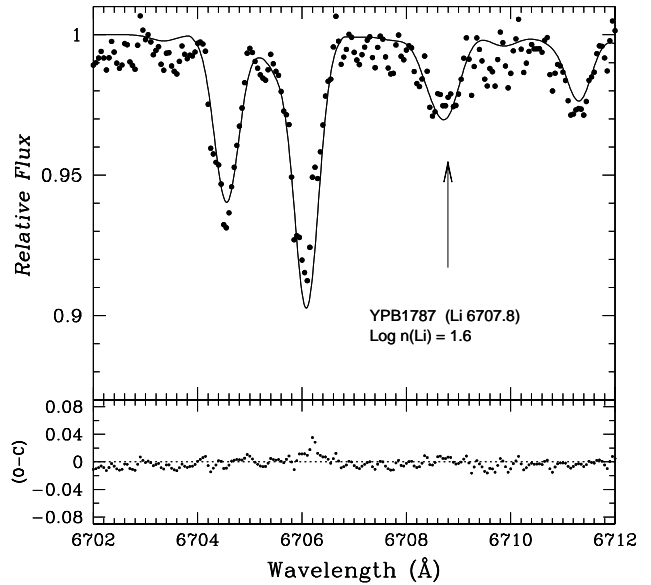}
\end{center}
\vspace{-0.2in}
\caption{Same as Figure \ref{fig:synt_obj637} for YPB1787} 
\label{fig:synt_obj1787}
\end{figure}

For our study, we used the effective temperature determined by the LDR method, which is believed to be more precise \citep{pasquini08}. For cluster M67 we adopted a metallicity of $[Fe/H] = 0.01 \pm 0.03$ \citep{randich06} and an age of 4.0 $\pm$ 0.5 Gyr \citep{dinescu95}. \citet{pasquini08} give a distance modulus for the cluster of $(m-M)_{0} = 9.63 \pm 0.06_{\rm stat} \pm 0.05_{\rm sys}$, which corresponds to a distance $d = 843^{+44}_{-41}$ pc. Intrinsic absolute magnitudes $M_{\rm V}$ were derived from this distance, the V magnitude given by \citet{pasquini08}, and a visual extinction $A_{\rm V} \sim 3.1 * E(B-V) = 0.127$ mag \citep{schultz&wiemer75} with $E(B-V) = 41 \pm 4$ mmag, the reddening of M67 \citep{taylor07}. We computed relative stellar luminosity compared to solar luminosity, using the bolometric corrections of \citet{flower96} and the associated error from the uncertainty in distance. 

The atmospheric parameters of all these stars are in excellent agreement with those of the Sun. However, the Li determination given in \citet{pasquini08} could be improved by spectral synthesis (see Sect. \ref{subsec:li_abundance}). The parameters of the solar analogs in M67 are summarized in Table \ref{tab:data_obs}.

\section{Observations and lithium abundance}
\label{sec:obs_lithium}

The spectra were acquired on three observation nights with the multi-object FLAMES/GIRAFFE spectrograph at the UT2/Kueyen ESO-VLT \citep{pasquini02} in MEDUSA mode\footnote{This is the observing mode in FLAMES in which 132 fibers feed the GIRAFFE spectrograph. Some fibers are set on the target stars and others on the sky background.} as part of a project searching for solar twins in M67 \citep{pasquini08}. The setting used was HR15N, which simultaneously covers the H$\alpha$ and \ion{Li}{i} resonance doublet at 670.8 nm with a resolution of R$\sim$17\,000. Three separate exposures were obtained to identify short and intermediate period binaries by comparing the radial velocities at different epochs. On average, combined spectra had a typical S/N of 80-110/pixel. For an extended description of sample selection, observations, and data reduction, see \citet{pasquini08}. 

\subsection{Lithium abundance}
\label{subsec:li_abundance}

The Li abundance of all M67 twins was derived from the \ion{Li}{i} resonance transition at $\lambda  6707~ \AA$ synthetic spectrum fitted to the GIRAFFE spectrum for the fundamental atmospheric parameters. First, we used \teff~ spectroscopic determination and surface gravity $\log g = 4.44$ dex, metallicity $[Fe/H] = 0.0$, projected surface velocity $v\sin i = 2.0~ {\rm km.s^{-1}}$, microturbulence velocity $\xi = 1.21~ {\rm km.s^{-1}}$ from \citet{pasquini08}. Atmospheric models were interpolated in the Kurucz grid \citep{kurucz93}, and the synthetic spectra were computed with MOOG \citep{sneden73}. The GIRAFFE instrumental broadening profile was set at $0.39~ \AA$, which is the FWHM of the GIRAFFE spectra. From the fundamental parameters, we derived $\log N{\rm (Li)} = 1.5$, $\log N{\rm (Li)} < 1.3$  and  $\log N{\rm (Li)} = 1.6$ respectively for YPB637, YPB1194, and YPB1787 (Fig. \ref{fig:synt_obj637}, Fig. \ref{fig:synt_obj1194}, and Fig. \ref{fig:synt_obj1787}), on the usual scale where $\log N{\rm (H)}$ = 12.00. 

Given the limited S/N and resolution of the YPB1194 spectrum, only an upper limit for the Li abundance of this star can be determined. The consequences are discussed  in Sect. \ref{sec:Results}. These values are higher than the determinations of \citet{pasquini08}, $\log N{\rm (Li)} = 1.4$ and $1.0$, and lower than $0.6$. Our Li abundance in the 10 solar twin candidates differed somewhat from the estimates obtained by \citet{pasquini08} because we used spectral synthesis instead of a calibration based on EWs, which may not be adequate for Li-depleted stars. For each star we determined $\log N{\rm (Li)}$ from synthetic models on the range of effective temperatures uncertainties. The procedure gives us an error of about 0.05 dex for $\log N{\rm (Li)}$ due to uncertainty in effective temperature. The errors arising from uncertainties surface gravity and microturbulent velocity are less than 0.01 dex, so can be ignored. The total error in $\log N{\rm (Li)}$ obtained by summing the error in model-fitting procedure \citep[see][]{bonifacio07} and effective temperature uncertainties in quadrature is equal to 0.10 dex as represented in Figs. \ref{fig:liteff_twins1194} and \ref{fig:liage_twins}.

\section{Stellar evolutionary models}
\label{sec:Models}

For the purposes of this study, stellar evolutionary models were computed using the Toulouse-Geneva stellar evolution code TGEC \citep{huibonhoa08}. Details on the physics of these models can be found in \citet{richard96,richard04}, \citet{donascimento00}, and \citet{huibonhoa08}. Standard input physics, non standard processes, diffusion, and rotation-induced mixing added in the models are described in \citet{donascimento09}. The method assumes the existence of an expected dependency between $\log N{\rm (Li)}$ and the star's mass and age for solar analogs stars \citep[e.g.,][]{montalban&schatzman00,charbonnel&talon05,xiong&deng09,donascimento09,baumann10}. This implies that stars with larger convection zones on the main sequence and a higher degree of differential rotation between the envelope and the radiative core present enhanced Li depletion, consequently we should know the angular momentum history of each studied star. Somehow, for solar twins stars  belonging to the same cluster, their angular momentum average may be approximated well by a mean angular momentum evolution history. In this context, our analysis is limited to solar-like stars on the main sequence and with a similar angular momentum history.

\subsection{Models and calibration}

We computed a grid of hundreds of evolutionary models of masses in the 0.90 to 1.10 \msun~ range with a step of 0.001 \msun, and with 
metallicities in the -0.030 to 0.050 range with a step of 0.010, from the zero-age main sequence (ZAMS) to the end of the hydrogen exhaustion in the core. Our evolutionary models were calibrated to match the observed solar effective temperature, luminosity, and Li abundance at the solar age. The calibration method of models is based on the \citet{richard96} prescription and is described in \citet{donascimento09}. In the following, we use models with a metallicity of $[Fe/H] = 0.01$, which is a simple average of different estimates of M67 metallicity \citep[see][]{pasquini08}.

\section{Results and discussion}
\label{sec:Results}

For each star we use an interpolation program to determine the mass. The uncertainties in $T_{\rm eff}^{\rm LDR}$ give rise to uncertainties in mass determination. The mass is interpolated from an HR diagram and from a \teff-$\log N{\rm (Li)}$ diagram. The results for each star are presented in Table \ref{tab:results}. Determination from the \teff-$\log N{\rm (Li)}$ diagram is more precise because Li abundance, more than luminosity, depends on mass \citep[see][]{donascimento09}. All stars of the sample have a mass within the range expected for the mass of a solar twin ($\pm$5 \% of the solar mass).

\begin{figure}[t!]
\begin{center}
\vspace{-0.1in}
\includegraphics[angle=0,height=9cm,width=9cm]{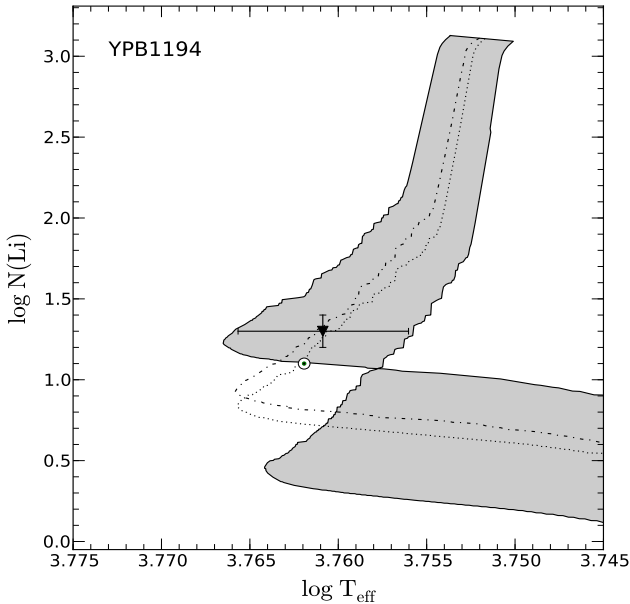}
\end{center}
\vspace{-0.2in}
\caption{Solar twin YPB1194 of open cluster M67 observed by \citet{pasquini08}: Li destruction along the evolutionary tracks as a function of effective temperature. The shaded zone represents the range of masses of TGEC models limited by the observational error bars. A model (dot-dashed lines) of 1.000 \msun \ passes through the observed point. The position of the Sun and the lithium destruction of a solar model (dotted line) are indicated. The Li abundance of the solar twin is an upper limit (triangle).} 
\label{fig:liteff_twins1194}
\end{figure}

In the following, we concentrate our analysis on objects YPB637, YPB1194, and YPB1787, which are the three most probable solar twins of the sample. They have a mass, an effective temperature, and Li abundance very close to the solar values. The temperature errors in the Table \ref{tab:results} are those given in Table A2 of \citet{pasquini08}, which are about 65K.

\begin{figure}[t!]
\begin{center}
\vspace{-0.1in}
\includegraphics[angle=0,height=9cm,width=9cm]{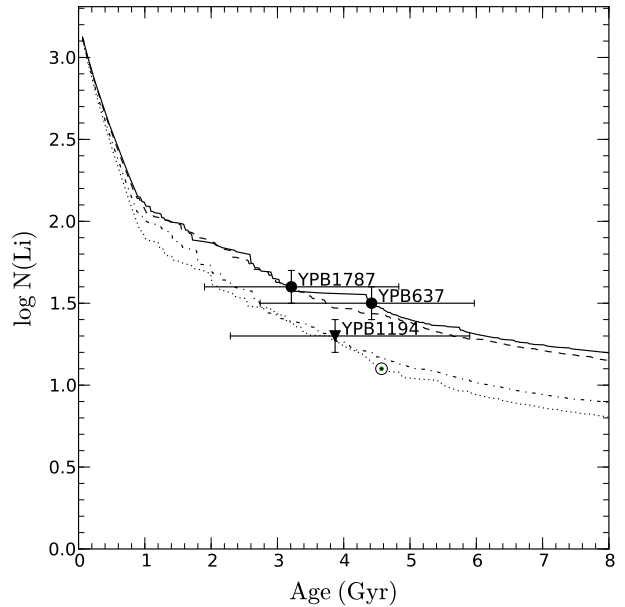}
\end{center}
\vspace{-0.2in}
\caption{Solar twins YPB637, YPB1194, and YPB1787 of open cluster M67 observed by \citet{pasquini08}: Li destruction along the evolutionary tracks as a function of age. The positions of the solar twins, the Sun, and Li destruction of a solar model (dotted line) are indicated. The Li abundance of the solar twin YPB1194 is an upper limit (triangle).} 
\label{fig:liage_twins}
\end{figure}

For these three stars, we calculated an evolutionary model of the mass determined by interpolation in the \teff-$\log N{\rm (Li)}$ diagram, and we determined the range of masses limited by the observational error in $T_{\rm eff}^{\rm LDR}$, which gives a value of $1.005^{+0.005}_{-0.005}$ \msun~ for the object YPB637, and $1.004^{+0.006}_{-0.008}$ \msun~ for the object YPB1787. For the object YPB1194, the determination of the Li abundance as an upper limit implies an upper limit for the mass determination: $1.000_{-0.008}$ \msun. In Figure \ref{fig:liteff_twins1194} we present the \teff-$\log N{\rm (Li)}$ diagram for YPB1194 as an example with the evolutionary tracks. The tracks of the evolution of the models pass through the observation point, as expected. The position of the observation point on the track and the tracks limited by the error give us an age of $4.42^{+1.55}_{-1.68}$ Gyr for YPB637 and $3.21^{+1.62}_{-1.31}$ Gyr for YPB1787. For YPB1194, the upper limit of the Li abundance gives a lower limit for the age: $3.87^{+2.03}$ Gyr. In Figure \ref{fig:liage_twins} we present Li destruction in the evolutionary model as a function of the age.

\subsection{Age determination of open cluster M67}

\begin{figure*}[t!]
\begin{center}
\vspace{-0.1in}
\includegraphics[angle=0,height=12cm,width=12cm]{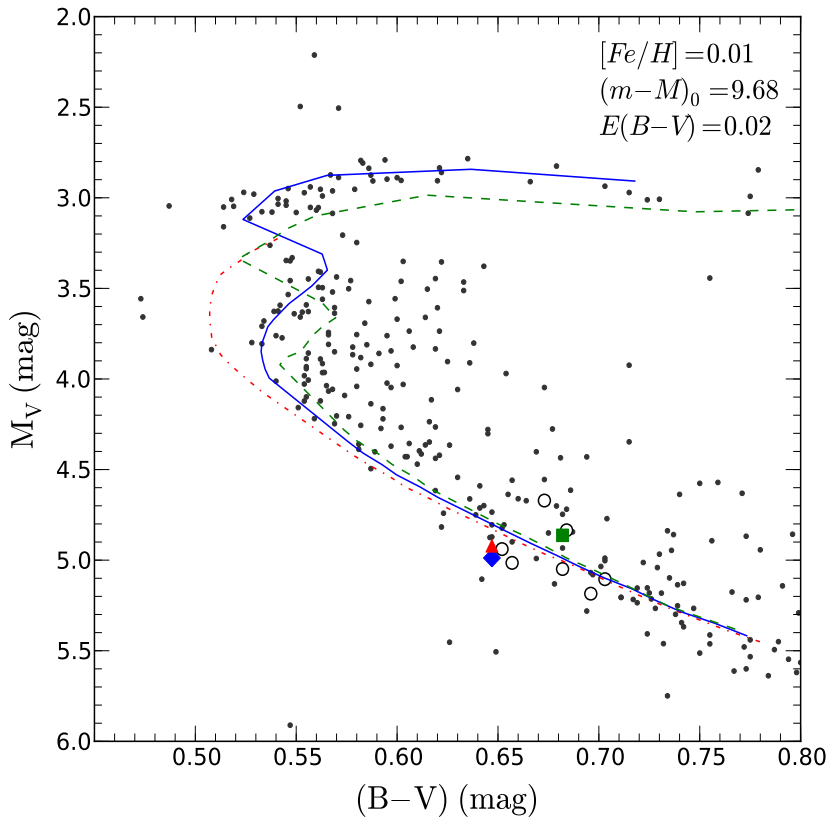}
\end{center}
\vspace{-0.2in}
\caption{Color-magnitude diagram of the solar analog sample of M67 from \citet{yadav08}. The dot-dashed isochrone, the continuous isochrone and the dashed isochrone correspond to an age of 3.21 Gyr, 3.87 Gyr, and 4.42 Gyr, respectively. The three best solar twins, YPB637, YPB1194, and YPB1787, are represented respectively as square, diamond, and triangle. The other solar analogs from \citet{pasquini08} are represented with open circles.} 
\label{fig:cmd-0.01}
\end{figure*}

The usual way to determine the age of an open cluster is to fit a theoretical isochrone with the observed main sequence and turn-off stars in a color-magnitude diagram. The position of the observed points in the diagram depends on the adopted interstellar reddening for the cluster, and the position of the isochrone varies according to metallicity and distance modulus. The ``hook'' of the turn-off is reproduced by calibrating the overshooting of the convective core. \citet{hobbs&thorburn91} determined an age of $5.2 \pm 1.0$ Gyr from the high-dispersion echelle spectra of five stars located along the main sequence of M67. In \citet{montgomery93}, the authors present a CCD photometry survey of M67. They derived a reddening $E(B-V) = 0.05 \pm 0.01$ and a metallicity $[Fe/H] = -0.05 \pm 0.03$. From cluster fitting to theoretical isochrones in a color-magnitude diagram, they found an age between 3 and 5 Gyr, but none of the isochrones fit the data consistently. \citet{dinescu95} used the sample of \citet{montgomery93}, cleaned up by cross-identification with the proper motion membership study of \citet{girard89}. They determined a distance modulus and the age of open cluster M67 by fitting the observed main sequence for single stars to the theoretical evolutionary tracks. They estimate an age of $4.0 \pm 0.5$ Gyr, and found that earlier models with overshoot of at most 0.1 $H_{\rm P}$ seem to reproduce the observations better. \citet{yadav08} find an age between 3.5 and 4.8 Gyr, by comparing their CMD, established from two-epoch archival observations with the Wide-Field Imager at the 2.2 m MPG/ESO telescope and reduced with new astrometric techniques, as described in \citet{anderson06}, to the following sets of theoretical isochrones: the BaSTI models provided by \citet{pietrinferni04} with $E(B-V) = 0.02$ and $[Fe/H] = 0.06$, the Padova models published by \citet{girardi00} with $E(B-V) = 0.02$ and $[Fe/H] = 0.00$, the Yonsei-Yale isochrones provided by \citet{yi01} in their second release with $E(B-V) = 0.02$ and $[Fe/H] = 0.00$, and the Victoria-Regina models published by \citet{vandenberg06} with $E(B-V) = 0.03$ and $[Fe/H] = -0.03$. They confirm that the convective core overshoot in M67 should be small, as pointed out by \citet{sandquist04} and \citet{vandenberg&stetson04}. \citet{vandenberg07} investigated the possibility of using the morphology of the M67 turn-off to put constraints on solar metallicity. Their models and isochrones computed on the assumption of a low-metal mix based on the solar abundances derived by \citet{asplund05}, and adopting a metallicity $[Fe/H] = 0.00$, a reddening $E(B-V) = 0.038$, and a distance modulus $(m-M)_{0} = 9.70$, do not predict the turnoff gap in the CMD. The same analysis using a 3.9 Gyr isochrone assuming the \citet{grevesse&sauval98} mix of heavy metals provides a good match to the morphology of the M67 CMD. \citet{magic10} confirm these results, but found that other physics in the models, e.g. element diffusion, nuclear reactions, prescription of core overshooting, also influence the stellar mass at which convective cores start to develop, and alter this result to the extent that isochrones constructed with models using low CNO solar abundances can also reproduce the turn-off morphology in M67. They find ages between 3.9 and 4.8 Gyr.

For each solar twin, we determined an age from the computed stellar model with mass previously determined. We found three different ages  in a narrow range of 1.2 Gyr. The age that best fits the color-magnitude diagram is 3.87 Gyr, the upper limit of the age of YPB1194, the best solar twin candidate (see Sect. \ref{subsec:cmd}). We used the two other age determinations to determine an error bar. Hence, our estimation for the age of open cluster M67 is $3.87^{+0.55}_{-0.66}$  Gyr, which is in very good agreement with the most recent results \citep{vandenberg07,yadav08,magic10}.

\subsection{The color-magnitude diagram}
\label{subsec:cmd}

Figure \ref{fig:cmd-0.01} presents the color-magnitude diagram of open cluster M67. The isochrones are constructed from computed models of metallicity $[Fe/H] = 0.01$ and masses from 0.90 to 1.31 \msun. The models of mass 1.19 \msun~ and higher are computed with a convective core overshooting parameter $\alpha_{ov}=0.01 H_{\rm P}$, whose occurrence is needed to provide a satisfactory match to the observed CMDs \citep[see][ and references therein]{kalirai01}. A very small convective core overshoot is needed to match the features near the main sequence turnoff region, in agreement with \citet{sandquist04}, \citet{vandenberg&stetson04}, and \citet{yadav08}. We used equation (3) and Table 4 of \citet{casagrande10} to compute  the $(B-V)$ color index of our models. We determined the absolute visual magnitude of the models from the luminosity using ($M_{bol}$)$\rm _{\odot}$ = 4.74 mag, and the bolometric corrections of \citet{flower96}. For the stars of the M67 sample, we used the $(B-V)$ and $V$ measurements of \citet{yadav08}. Our best fit of the main sequence and the ``hook'' of the turn-off is achieved with isochrone 3.87 Gyr, using a distance modulus $(m-M)_{0} = 9.68$ and a reddening $E(B-V) = 0.02$ mag. The value of the distance modulus is in good agreement with former determinations \citep{vandenberg07,yadav08,pasquini08}. The reddening is lower than the value obtained by \citet{taylor07} but, as pointed out by \citet{yadav08}, given the intrinsic uncertainties in calibration of $E(B-V)$, the true reddening could be anywhere between 0.02 and 0.04 mag.

\section{Conclusions}

\label{sec:Conclusions}

In this paper, we applied the method developed by \citet{donascimento09} to the solar analogs of open cluster M67 identified by \citet{pasquini08}, in order to determine their masses as accurately as possible. We identified three objects, YPB637, YPB1194, and YPB1787, whose effective temperature, luminosity, mass, and Li abundance values are very close to the solar ones. For these three solar twins we derived ages of $4.42^{+1.55}_{-1.68}$ Gyr for YPB637, and $3.21^{+1.62}_{-1.31}$ Gyr for YPB1787. For YPB1194, the lower limit of the age is $3.87^{+2.03}$ Gyr. These estimations allowed us to construct isochrones on a color-magnitude diagram for open cluster M67, using the sample of \citet{yadav08}. The isochrone 3.87 Gyr matches both the main-sequence and the turn-off regions, using a distance modulus $(m-M)_{0} = 9.68$ and a reddening $E(B-V) = 0.02$ mag. The three solar twins allowed us to estimate the age of the open cluster as $3.87^{+0.55}_{-0.66}$ Gyr. We thus confirm the importance of M67 as a cluster with a slightly lower age than the solar one, and the mass of the solar twin candidates is also very similar to solar; therefore, M67 is an important cluster when searching for solar twins.

\begin{acknowledgements}
This research used SIMBAD and VIZIER databases, operated at CDS (Strasbourg, France). JRM thanks the support provided by FCT (Ciencia 
2007). Research activities of the Stellar Board at the Federal University of Rio Grande do Norte are supported by continuous grants from CNPq and FAPERN Brazilian Agencies. The authors are grateful for the support from FCT/CAPES cooperation agreement n$^{o}$237/09. M. Castro thanks C. Cort\'es for the interpolation program.
\end{acknowledgements}

\bibliographystyle{aa}

{}


\begin{table*}
\begin{threeparttable}
\caption{Parameters of the observed solar analogues in M67 \tnote{1}.}
\label{tab:data_obs}
\centering
\begin{minipage}[t]{\textwidth}
\centering
\begin{tabular}{cccrr}
\hline
Object & $T_{\rm eff}^{\rm LDR}$ & $(L/L_{\odot})$ & $\log N{\rm (Li)}^{\rm NLTE}$ & $\log N{\rm (Li)}_{moog}\tnote{*}$ \\
 & (K) & & & \\
\hline
285  & 5836 $\pm$ 67 & 1.112$^{+0.119}_{-0.106}$ & $<$0.6 &     0.6 \\
637  & 5806 $\pm$ 65 & 1.089$^{+0.117}_{-0.103}$ &    1.4 &     1.5 \\
1101 & 5756 $\pm$ 60 & 0.925$^{+0.099}_{-0.088}$ & $<$0.6 & $<$ 0.6 \\
1194 & 5766 $\pm$ 64 & 0.976$^{+0.105}_{-0.093}$ & $<$0.6 & $<$ 1.3 \\
1303 & 5716 $\pm$ 64 & 0.960$^{+0.103}_{-0.091}$ & $<$0.6 &     1.2 \\
1304 & 5704 $\pm$ 64 & 0.886$^{+0.095}_{-0.084}$ & $<$0.5 & $<$ 0.8 \\ 
1315 & 5874 $\pm$ 58 & 1.286$^{+0.138}_{-0.122}$ &    1.4 &     1.8 \\
1392 & 5716 $\pm$ 63 & 0.821$^{+0.088}_{-0.078}$ & $<$0.6 & $<$ 0.6 \\
1787 & 5768 $\pm$ 70 & 1.039$^{+0.111}_{-0.099}$ &    1.0 &     1.6 \\
2018 & 5693 $\pm$ 74 & 1.035$^{+0.111}_{-0.098}$ & $<$0.5 & $<$ 1.3 \\
\hline
\end{tabular}
\begin{tablenotes}
\footnotesize{
\item[1] Note: Cols 1-4 from Pasquini et al. (2008).
\item[*] Our determinations of the Li abundances using the MOOG program
}
\end{tablenotes}

\end{minipage}
\end{threeparttable}
\end{table*}

\begin{table*}
\caption{Results of mass determination from interpolation in the $T_{\rm eff}$-$\log N{\rm (Li)}$ diagram.}
\label{tab:results}
\centering
\begin{minipage}[t]{\textwidth}
\centering
\begin{tabular}{ccc}
\hline
Object &   Mass \\
 &  (\msun) \\
\hline
 285 &  1.001$^{+0.030}_{-0.011}$ \\
 637 & 1.005$^{+0.012}_{-0.005}$ \\
1101 &  0.988$^{+0.008}_{-0.008}$ \\
1194 &  0.999$^{+0.007}_{-0.008}$ \\
1303 &  0.985$^{+0.006}_{-0.009}$ \\
1304 &  0.983$^{+0.005}_{-0.009}$ \\
1315 &  1.018$^{+0.022}_{-0.013}$ \\
1392 &  0.985$^{+0.006}_{-0.009}$ \\
1787 &  1.004$^{+0.007}_{-0.008}$ \\
2018 &  0.982$^{+0.006}_{-0.009}$ \\
\hline
\end{tabular}
\end{minipage}
\end{table*}


\end{document}